# An Ultra-Low Profile Metasurface for Enhanced Non-Diffractive OAM Beam Generation Based on Sparse Feed Array

Jinyang Bi
*School of Communication Engineering*
*Xidian University*
Xi'an, China
jybi@stu.xidian.edu.cn

Fan Qin
*School of Communication Engineering*
*Xidian University*
Xi'an, China
fqin@xidian.edu.cn

Chao Gu
*the ECIT Institute*
*Queen's University Belfast*
Belfast, U.K
chao.gu@qub.ac.uk

Hailin Zhang
*School of Communication Engineering*
*Xidian University*
Xi'an, China
hlzhang@xidian.edu.cn

Wenchi Cheng
*School of Communication Engineering*
*Xidian University*
Xi'an, China
wccheng@xidian.edu.cn

***Abstract*—This paper proposes a novel ultra-low profile transmitted metasurface to generate enhanced non-diffractive orbital angular momentum (OAM) beams, employing a sparse feed array (SFA) to create a quasi-plane wave excitation for the first time. Our simulation indicates that with uniform amplitude excitation, the non-diffraction performance of Bessel beam produced by metasurface, surpasses that of conventional single-feed excitation. Based on this principle, a 5 × 5 sparse feed array is introduced and positioned less than one wavelength from the metasurface, ensuring a quasi-uniform amplitude excitation across all units. Further, this metasurface leverages its flexible phase control capability and integrates the spatial phase, OAM phase, and axicon phase to generate an enhanced high-order Bessel beam. The simulated results confirm a successful non-diffractive Bessel beam generation carrying OAM with mode *l* = +2, exhibiting reduced beam divergence and higher gain. This design also offers benefits of ultra-low profile, high aperture efficiency, low structural complexity.**

***Keywords—high-order Bessel beam, non-diffraction, orbital angular momentum (OAM), metasurface, sparse array.***

## I. INTRODUCTION

RECENTLY, the existing communication technology has modulated signals from time, space, frequency, and polarization dimensions to enhance capacity. However, the finite spectral and polarization resources impose fundamental constraints. The mode-division multiplexing (MDM), based on orbital angular momentum (OAM), has emerged as a promising solution, as OAM eigenstates offer theoretically infinite orthogonal modes, enabling numerous independent channels. Compared to conventional multiplexing schemes, this approach significantly enhances channel capacity and spectral utilization efficiency.

However, the OAM wave suffers from beam divergence during propagation. As one of the classical non-diffractive beams, Bessel beam, features transversal Bessel-function intensity field distribution and diffraction-resisting property [1]. High order Bessel beams, integrating OAM with Bessel beam property, not only encode OAM information but also mitigate the vortex wave divergence. Some high order Bessel beams based on radial line slotted array (RLSA) [2], reflect- or transmit-arrays [3]-[7], and 3-dimensional printed lenses [8] have been reported. Further, to create ideal Bessel beam, an amplitude-phase-modulated surface (APMS) has been utilized, ensuring the excitation of each unit following the nth-order Bessel function of the first kind [9]. However, this design introduces extra structural complexity and exhibits low-efficiency property, limiting its practical application. Thus, Maximizing the non-diffraction of Bessel OAM beam while maintaining a limited aperture, low complexity, and high gain and efficiency still remains a challenge.

To create Bessel beams using metasurface, our simulation indicates that under uniform amplitude excitation, the non-diffraction performance of Bessel OAM beams surpasses that of single-feed excitation. To address single-feed limitation and verify our simulation, this paper presents an ultra-low profile metasurface excited by a sparse feed array (SFA), introduced to provide a quasi-uniform amplitude excitation. Based on the SFA, this metasurface achieves high-order Bessel beam carrying $l = +2$ OAM mode with enhanced non-diffraction performance and higher gain.

## II. WORKING PRINCIPLE

Configurations of the transmitted metasurface antennas for various OAM beam generation are depicted in Fig. 1. The metasurfaces M1 and M2, illuminated by a single feeding, implement the conventional diffractive OAM generation and high-order Bessel OAM beam generation above the surface aperture, respectively. For M1, the required compensation phase of the *mn*th element can be calculated by

$$\phi_{mn} = k_0 \left| \vec{r}_{mn} - \vec{r_f} \right| - l\varphi_{mn} \qquad (1)$$

where $r_{mn}$ and $\varphi_{mn}$ is the position vector and azimuth angle of the *mn*th element. $r_f$ is the position vector of the feed source. $l$ is the desired OAM mode. The first term in right-hand side in (1) is to mimic the functionality of the focusing hyperbolic lens to convert the spherical phase front from the feed source to plane wave, while the second term emulates the spiral phase plate (SPP) to generate the vortex wave.

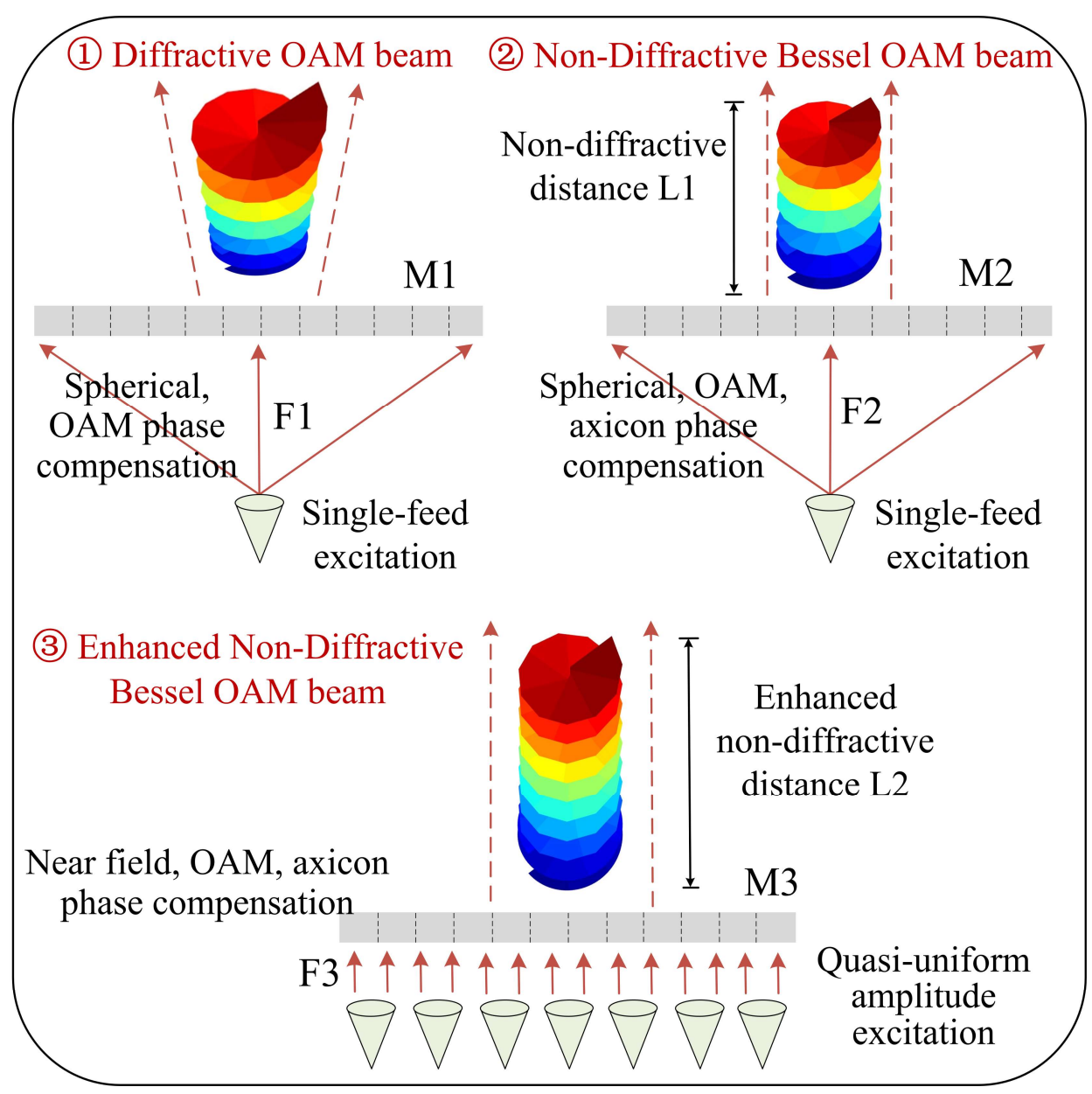

Fig. 1. Configurations of the transmitted metasurface antennas for OAM beam generation. (a) M1 for conventional OAM beam generation (b) M2 for high-order Bessel beam generation. (c) Ultra-low profile M3 for enhanced high-order Bessel beam generation.

To generate high order Bessel beam carrying OAM, the required compensation phase distribution of M2 should be

$$\phi_{mn} = k_0 \left| \vec{r}_{mn} - \vec{r}_f \right| - l\varphi_{mn} + k_0 \left| \vec{r}_{mn} \right| \sin\delta \qquad (2)$$

where δ is the axicon angle of the Bessel beam, satisfying the following relationship:

$$\tan\delta = \frac{k_\rho}{k_z} \qquad (3)$$

This phase combines spatial phase, OAM phase, and axicon phase using one single lens. After tradeoff between aperture size and non-diffractive range, the axicon angle of M2 is set as δ = 10° after optimization. Fig. 2 shows the calculated compensation phase distribution of M1 and M2.

For M2, the single-feed excitation results in strong center excitation energy and weak edge excitation energy, deviating from ideal Bessel beam generation conditions. To enhance Bessel beam performance, a sparse feed array is introduced to realize a uniform-amplitude excited metasurface M3, as depicted in Fig. 1(c). A comparison is performed between M2 and M3 to evaluate the impact of various feeding methods on non-diffraction performance of Bessel beam.

Fig. 3 illustrates the electric field distributions of M1, M2, and M3 generating conventional OAM beam, high-order Bessel beam, and enhanced Bessel beam with $l$ = +2 mode on the longitudinal plane (*yoz* plane) respectively, simulated by MATLAB software. Both M1, M2, and M3 are composed of 50 × 50 units, the period of which is set as $P$ = 10 mm, corresponding to 0.34 λ, where λ is the free space wavelength at center frequency of 10.25 GHz. The observational region is 50 - 3000 mm away from the metasurface. The theoretical maximum non-diffractive distance is 1400 mm. As shown in Fig. 3, the E-field radiated from M1 spreads out significantly along the propagation direction. In contrast, the E-field from M2 is concentrated around the central axis over a long distance of around 1400 mm. Since M3 employs a uniform amplitude excitation, its radiated energy is significantly stronger than M2 within the non-diffraction range, while the beams on both sides of the axis remain more parallel. Beyond the non-diffraction range, the divergence angle of OAM beam in M3 is also significantly smaller than M2. Thus, the M3, under uniform amplitude excitation, demonstrates an extended non-diffraction performance compared to M2.

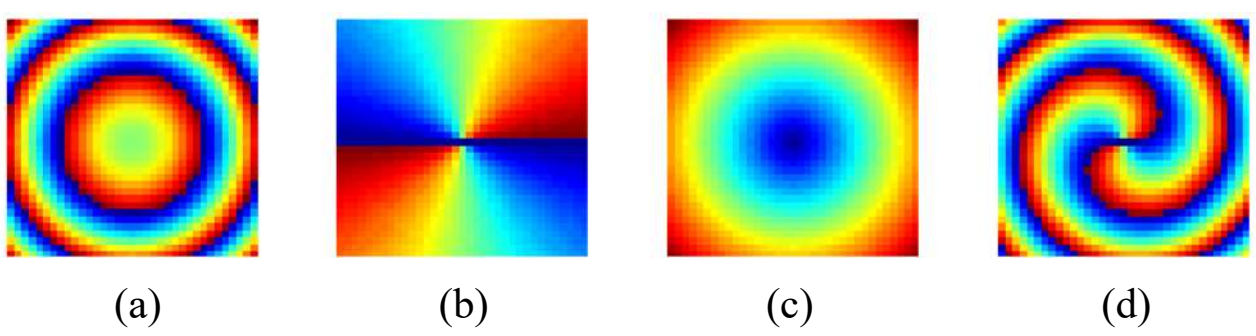

Fig. 2. Compensation phase distributions of M1 and M2 (a) spherical phase. (b) OAM phase. (c) axicon phase. (d) final Bessel beam compensation.

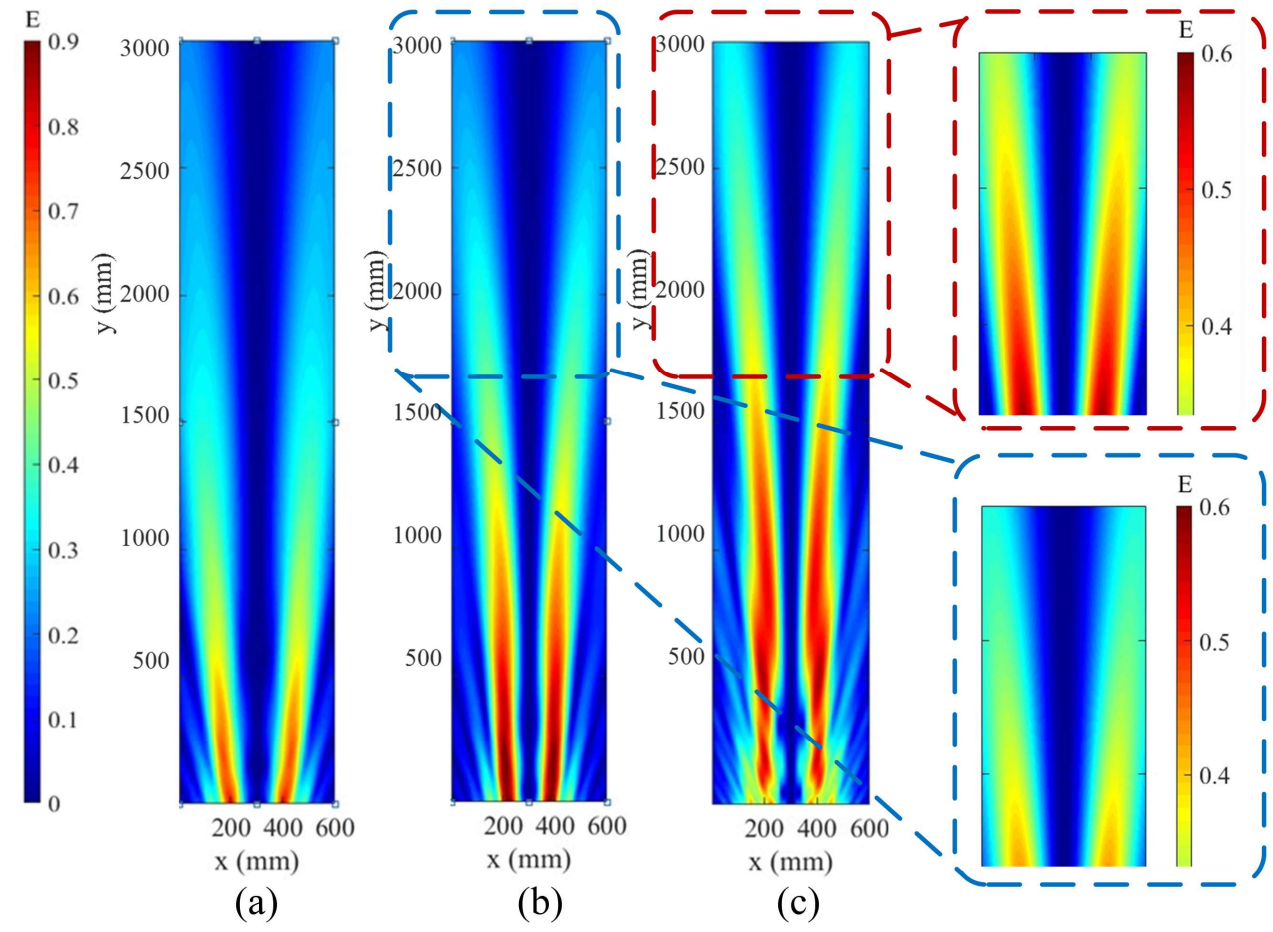

Fig. 3. Normalized E-field distributions of M1, M2, and M3 for generating OAM beams on longitudinal plane. (a) M1 for conventional OAM beam (b) M2 for high-order Bessel beam. (c) M3 for enhanced Bessel beam.

## III. Design of The Proposed Metasurface

To further demonstrate the enhancement of Bessel OAM beam non-diffraction with uniform amplitude excitation, the proposed metasurface fed by sparse feed array was designed and simulated in HFSS software, forming the M3.

### *A. Unit Cell Configuration*

The unit cell structure is given in Fig. 4, which integrates a metallic double arrow-shaped resonator in the middle layer sandwiched by two metal orthogonal grids printed on the substrate, arranging parallel to x- and y-axis, respectively. The polarization grid has polarization selectivity, permitting only the perpendicular polarized component of incident wave to propagate through while reflecting the parallel polarized component. The middle patch is rotated by 45° relative to the x-axis, enabling to rotate the transmitted wave polarization by 90°. Thus, with the incident wave illuminated from top to bottom on the unit, it converts the y-polarized wave into a x-polarized wave while the entire x-polarized wave is reflected.

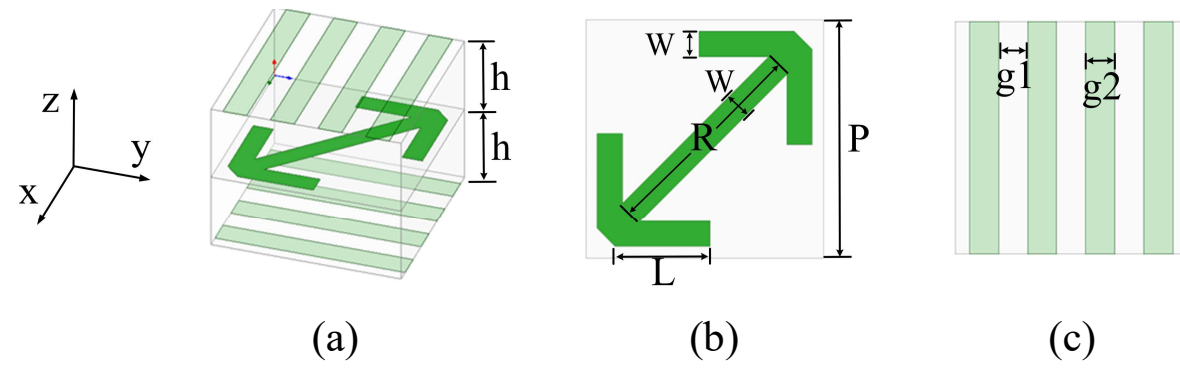

Fig. 4. Configuration of the transmitted metasurface unit cell. (a) 3-D view. (b) metal pattern in middle layer. (c) polarization grid layer. The geometrical parameters are: $h$ = 3 mm, $P$ = 7.5 mm, $W$ = 0.8 mm, $R$ = 6.5 mm, $g1$ = 0.9375 mm, and $g2$ = 0.9375 mm.

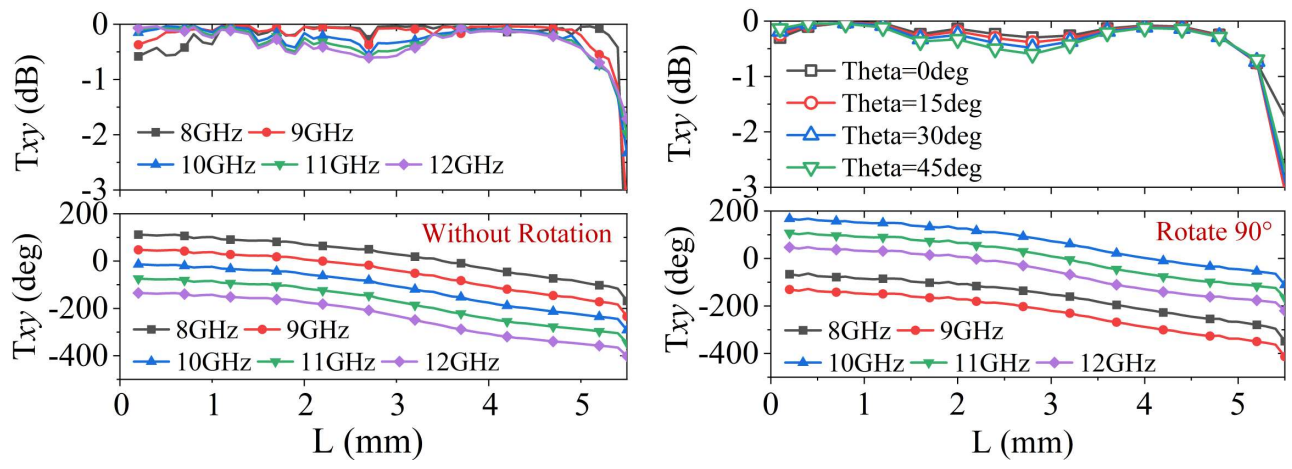

Fig. 5. Transmission magnitude and phase versus parameter "$L$" at different frequencies and oblique incident degrees.

F4B is adopted as the dielectric substrate, with a relative permittivity of 2.65, loss tangent of 0.001, and thickness $h$ of 3 mm. The total thickness of the unit is 6 mm, corresponding to 0.2 λ, while the period of the unit is selected as 7.5 mm, corresponding to 0.26 λ of 10.25 GHz.

The transmission coefficients of the unit for amplitude and phase from 8.0 to 12.0 GHz are shown in Fig. 5. By varying "$L$", this unit exhibits phase changing range exceeding 180°, while maintaining a transmission coefficient below -0.6 dB across entire frequency bands. Further, by rotating the middle patch by 90°, the whole phase-tunning range exceeding 360°. The phase curves are almost parallel. Further, under oblique incidences, only minor amplitude deviations occur, revealing that the performance of the unit remains stable.

### *B. Sparse Feed Array Design*

Fig. 6 depicts a microstrip parasitic patch antenna, which consists of two dielectric substrates of F4B with thickness of 0.8 mm, a metal ground, a radiation patch with $R1$ = 7.8 mm, and a parasitic patch with $R2$ = 9.4 mm. The simulated peak gain is 9.76 dBi at 10.25 GHz, as shown in Fig. 6(d).

Based on this single antenna, a designed 5 × 5 sparse feed array and its feed network are shown in Figs. 7. This array element spacing is set to 52.5 mm, corresponding to 1.8 λ. The feed network enables equal amplitude and phase feeding of each array unit through an impedance matching of power divider, as shown in Fig. 8. Based on this SFA, the electric field and phase distribution of a receiving surface with 262.5 mm × 262.5 mm aperture, positioned above and less than one wavelength (22 mm, corresponding to 0.75 λ) from the SFA, are shown in Figs. 7(c) and (d). The electric field distribution is more uniform than that of a spherical wave on the same surface and closely approximates a quasi-uniform amplitude.

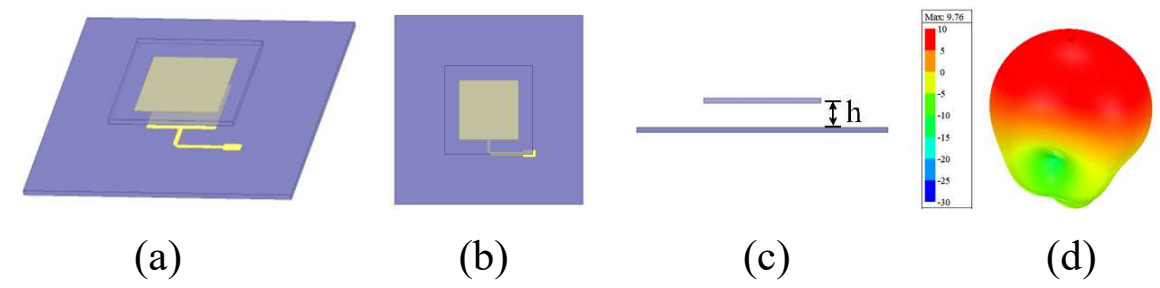

Fig. 6. Configuration and radaition pattern of the microstrip parasitic antenna. (a) 3-D view. (b) top view. (c) front view. (d) radiation pattern.

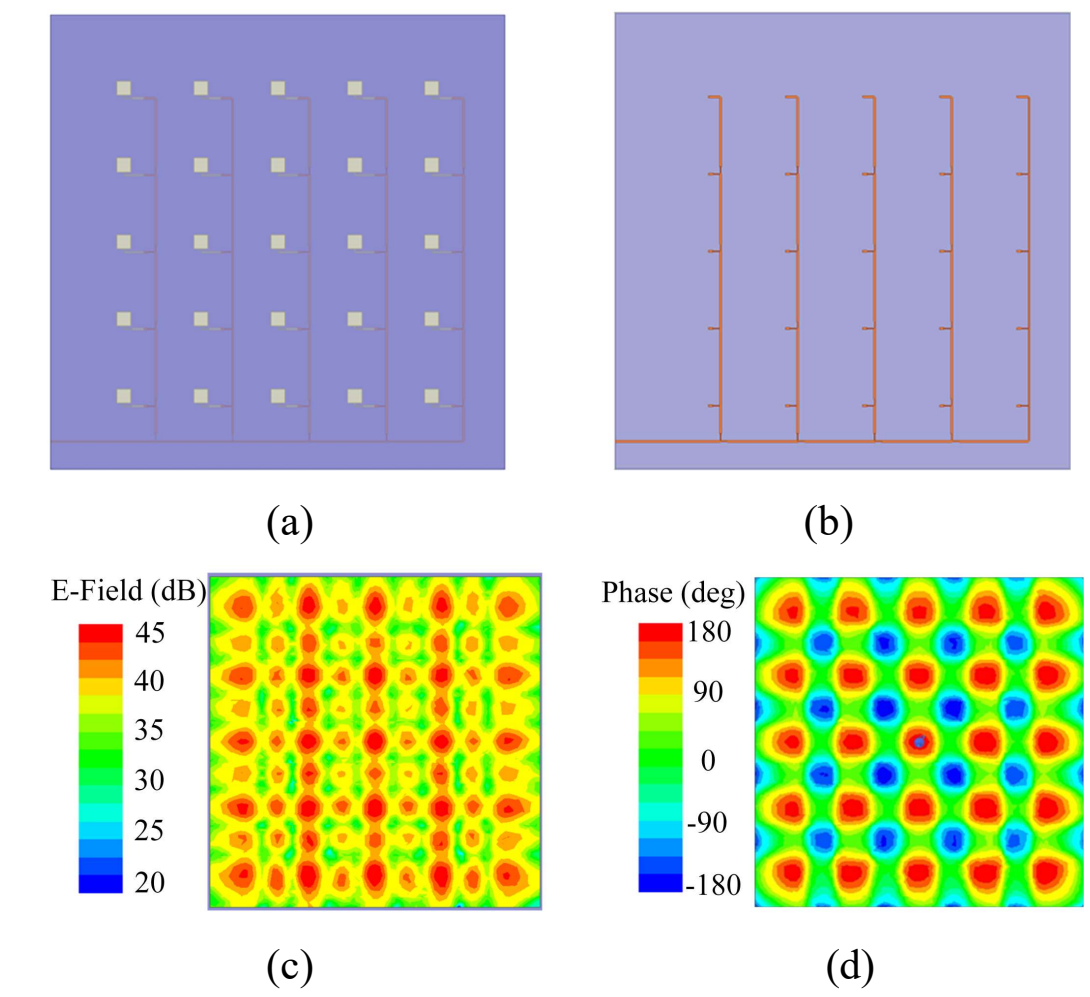

Fig. 7. (a) Configuration (b) feed network of the SFA; (c) E-field and (d) phase distribution of the receiving surface above the SFA.

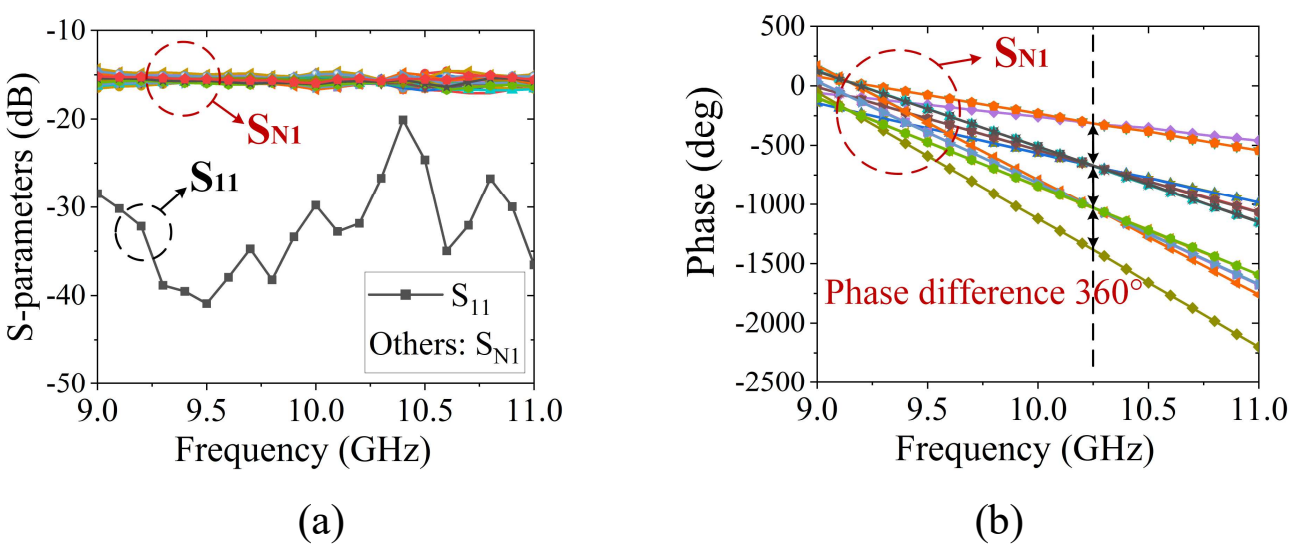

Fig. 8. (a) S-parameters (b) phase response of the power divider of SFA.

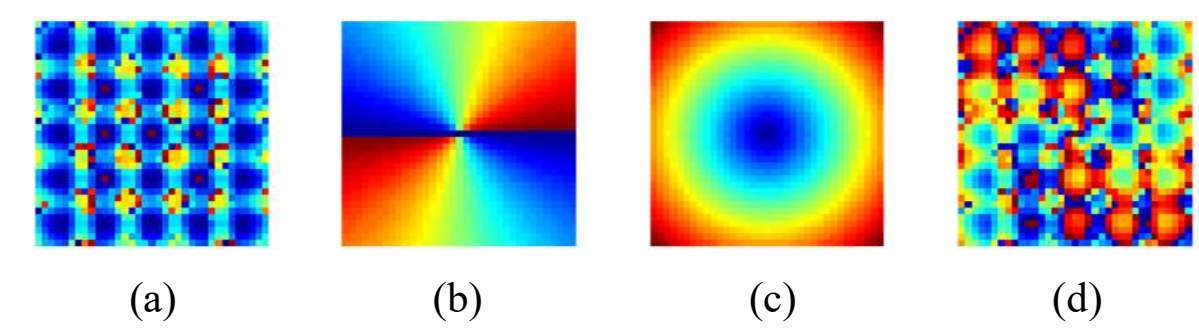

Fig. 9. Compensation phase distributions of M1 and M2 (a) spherical phase. (b) OAM phase. (c) axicon phase. (d) final Bessel beam compensation.

### *C. Design of the Metasurface*

Since M3 is located in the near-field region of the SFA, expressing the spatial phase compensation mathematically is challenging. By obtaining the phase distribution of receiving surface from HFSS simulations, as shown in Fig. 9(a), the corresponding phase is derived to substitute for the spherical wave phase compensation. Thus, based on (2), the final phase compensation for generating a Bessel beam carrying $l$ = +2 OAM mode by M3 under quasi-uniform amplitude excitation

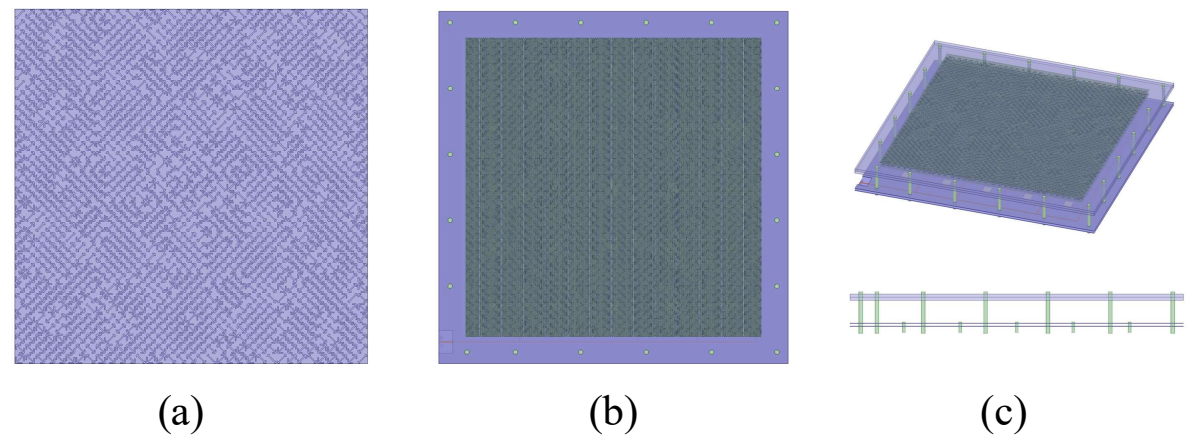

Fig. 10. (a) the middle layer (b) top view (c) 3-D view (d) side view of M3.

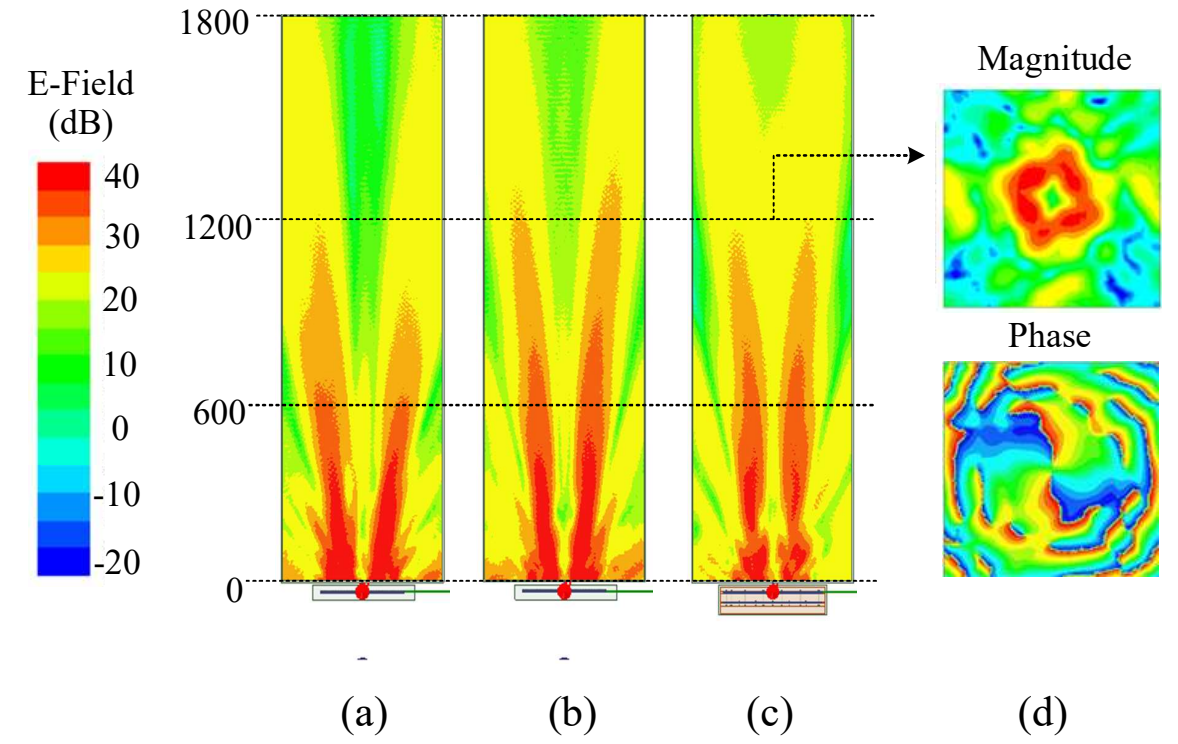

Fig. 11. (a) Non-diffraction performance of (b) M1 for conventional OAM beam generation. (b) M2 for high-order Bessel beam generation. (c) M3 with quasi-uniform amplitude excitation for high-order Bessel beam generation.

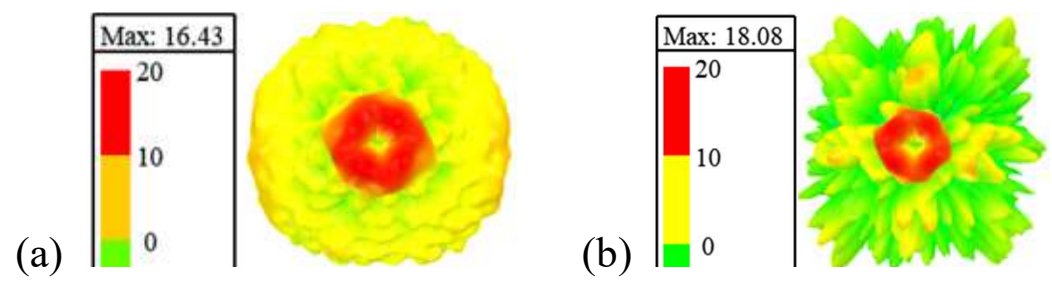

Fig. 12. (a) conventional Bessel beam with $l$ = +2 OAM mode by M2. (b) enhanced Bessel beam with $l$ = +2 OAM mode by M3.

is presented in Fig. 9(d), and its corresponding middle pattern layer is depicted in Fig. 10(a).

By combining SFA with metasurface M3 and mounting with nylon screws, the ultra-low profile metasurface antenna excited by quasi-uniform amplitude is finally obtained and depicted in Figs. 10(b) and (c). The aperture of the antenna is 262.5 mm × 262.5 mm (8.96 λ × 8.96 λ). The focal length of F is optimized to 22 mm, while the F/D ratio surpass 0.1.

## IV. SIMULATION RESULTS

To validate the design strategy, both metasurfaces M1, M2 and M3 with various feed sources carrying $l$ = +2 OAM mode are simulated in HFSS, while electric fields on longitudinal plane are plotted in Figs. 11(a)-(c). The E-field radiated from M1 spreads out significantly along the propagation direction, whereas the E-field from M2 exhibits a noticeable reduction in divergence angle. However, due to the finite aperture, it does not fully achieve an ideal non-diffraction beam. For M3, within non-diffraction range, the divergence angle of Bessel beam is further reduced, effectively maintaining parallel. Beyond non-diffraction range, energy of M3 remains more concentrated near the axis, with significantly suppressed divergence. Thus, with limited aperture, a uniform amplitude excitation of metasurface shows an extended non-diffraction performance compared to single-feed excitation.

Further the magnitude and phase distributions of receiving surface positioned 1200 mm above M3 are shown in Fig. 11 (d), exhibiting a hollow energy and spiral phase distribution. The radiation patterns of M2 and M3 are shown in Fig. 12. Compared with M2, due to the enhanced radiation effect of M3's edge elements, the radiation pattern of M3, from top view, expands from a circular shape to a square. Under the same aperture, M2 achieves a peak gain of 16.4 dBi, while M3 reaches 18.1 dBi. The aperture efficiency of M3 is 6.4%, which is higher than 4.3% of M2. Therefore, the uniform-amplitude excited metasurface design also offers advantages of high gain and high aperture efficiency.

## V. CONCLUSION

This paper presents a novel ultra-low profile metasurface with a quasi-uniform amplitude excitation by integrating a sparse feed array configuration. Compared with conventional single-feed metasurface, this design implements high-order Bessel beam carrying OAM with enhanced non-diffraction performance. This antenna also shows the benefits of high gain, high aperture efficiency, and low structural complexity.